%% file: spatial-exploration.tex
\documentclass[sigconf]{acmart}

\AtBeginDocument{%
  \providecommand\BibTeX{{%
    \normalfont B\kern-0.5em{\scshape i\kern-0.25em b}\kern-0.8em\TeX}}}

\usepackage{enumerate}
\usepackage{enumitem}
\usepackage{multirow}
\usepackage{graphicx}
\usepackage{caption}
\usepackage{color,soul}
\usepackage{subcaption}
\usepackage{fontawesome5}
\usepackage{tabularray}
\usepackage{colortbl}
\usepackage{acmart-taps}

\definecolor{pumpkin}{RGB}{211, 84, 0}
\definecolor{red}{RGB}{255, 0, 0}

\copyrightyear{2024}
\acmYear{2024}
\setcopyright{acmlicensed}\acmConference[CHI '24]{Proceedings of the CHI Conference on Human Factors in Computing Systems}{May 11--16, 2024}{Honolulu, HI, USA}
\acmBooktitle{Proceedings of the CHI Conference on Human Factors in Computing Systems (CHI '24), May 11--16, 2024, Honolulu, HI, USA}
\acmDOI{10.1145/3613904.3642632}
\acmISBN{979-8-4007-0330-0/24/05}

\begin{document}

\title[SPICA: Interactive Video Content Exploration through Augmented Audio Descriptions for BLV Viewers]{SPICA: Interactive Video Content Exploration through Augmented Audio Descriptions for Blind or Low-Vision Viewers}

\author{Zheng Ning}
\email{zning@nd.edu}
\affiliation{%
  \institution{University of Notre Dame}
  \city{Notre Dame}
  \state{Indiana}
  \country{USA}
}

\author{Brianna L. Wimer}
\email{bwimer@nd.edu}
\affiliation{%
  \institution{University of Notre Dame}
  \city{Notre Dame}
  \state{Indiana}
  \country{USA}
}

\author{Kaiwen Jiang}
\email{k1jiang@ucsd.edu}
\affiliation{%
  \institution{University of California San Diego}
  \city{La Jolla}
  \state{California}
  \country{USA}
}

\author{Keyi Chen}
\email{kec020@ucsd.edu}
\affiliation{%
  \institution{University of California San Diego}
  \city{La Jolla}
  \state{California}
  \country{USA}
}

\author{Jerrick Ban}
\email{jban@nd.edu}
\affiliation{%
  \institution{University of Notre Dame}
  \city{Notre Dame}
  \state{Indiana}
  \country{USA}
}

\author{Yapeng Tian}
\email{yapeng.tian@utdallas.edu}
\affiliation{%
  \institution{University of Texas at Dallas}
  \city{Richardson}
  \state{Texas}
  \country{USA}}

\author{Yuhang Zhao}
\email{yuhang.zhao@cs.wisc.edu}
\affiliation{%
  \institution{University of Wisconsin-Madison}
  \city{Madison}
  \state{Wisconsin}
  \country{USA}}

\author{Toby Jia-Jun Li}
\email{toby.j.li@nd.edu}
\affiliation{%
  \institution{University of Notre Dame}
  \city{Notre Dame}
  \state{Indiana}
  \country{USA}}

\begin{abstract}
Blind or Low-Vision (BLV) users often rely on audio descriptions (AD) to access video content. However, conventional static ADs can leave out detailed information in videos, impose a high mental load, neglect the diverse needs and preferences of BLV users, and lack immersion. To tackle these challenges, we introduce \textsc{Spica}, an AI-powered system that enables BLV users to interactively explore video content. Informed by prior empirical studies on BLV video consumption, \textsc{Spica} offers interactive mechanisms for supporting temporal navigation of frame captions and spatial exploration of objects within key frames. Leveraging an audio-visual machine learning pipeline, \textsc{Spica} augments existing ADs by adding interactivity, spatial sound effects, and individual object descriptions without requiring additional human annotation. Through a user study with 14 BLV participants, we evaluated the usability and usefulness of \textsc{Spica} and explored user behaviors, preferences, and mental models when interacting with augmented ADs.
\end{abstract}



\renewcommand{\shortauthors}{Zheng Ning, et al.}

\sloppy
\maketitle

\input{1-Introduction}
\input{2-Related_work}
\input{3-System}
\input{4-ML_pipeline}

\input{5-User_study}
\input{6-DiscussionFutureWorkConclusion}

\begin{acks}
This work was supported in part by an AnalytiXIN Faculty Fellowship, an NVIDIA Academic Hardware Grant, a Google Cloud Research Credit Award, a Google Research Scholar Award, University of Wisconsin—Madison Office of the Vice Chancellor for Research and Graduate Education with funding from the Wisconsin Alumni Research Foundation, and NSF Grants 2211428 and 2326378. Yapeng Tian was supported by a gift from Cisco systems. Any opinions, findings or recommendations expressed here are those of the authors and do not necessarily reflect views of the sponsors. We would like to thank Martez Mott for the useful discussion on qualitative analysis; Gaurav Jain and Sitong Wang for the assistance in user study.
\end{acks}

\balance
\bibliographystyle{ACM-Reference-Format}
\bibliography{spatial-exploration}

\clearpage

\onecolumn

\appendix
\section{Participant demographics}
\aptLtoX[graphic=no,type=html]{\begin{table}
\centering
\begin{tabular}{l|l|l|l|l}
\toprule
\textbf{Participant ID} & \textbf{Age} & \textbf{Gender} & \textbf{Onset} & \textbf{Level of Visual Impairment}        \\
\midrule
P0 (Pilot study)                     & 67           & Female          & Acquired       & Legal blindness                            \\
P1                      & 24           & Female          & Acquired       & Legal blindness                            \\
P2                      & 30           & Female          & Congenital     & Total Blindness                            \\
P3                      & 23           & Female          & Congenital     & Total Blindness                            \\
P4                      & 28           & Female          & Congenital     & Blindness with some light/color perception \\
P5                      & 23           & Male            & Acquired       & Blindness with some light/color perception \\
P6                      & 44           & Female          & Congenital     & Legal blindness                            \\
P7                      & 24           & Male            & Acquired       & Blindness with some light/color perception \\
P8                      & 18           & Male            & Congenital     & Total Blindness                            \\
P9                     & 41           & Male            & Acquired       & Blindness with some light/color perception \\
P10                     & 23           & Male            & Congenital     & Total Blindness                            \\
P11                     & 32           & Male            & Congenital     & Total Blindness                            \\
P12                     & 33           & Female          & Congenital     & Total Blindness                            \\
P13                     & 38           & Female          & Congenital     & Total Blindness                            \\
P14                     & 33           & Male            & Acquired       & Legal blindness                           \\
\hline 
\end{tabular}
\caption{Participant demographics for our user study}
\label{tab:participants}
\end{table}}
{\begin{table*}[!hb]
\centering
\begin{tblr}{
  row{1} = {c},
  vline{2-5} = {-}{},
  hline{1,17} = {-}{0.08em},
  hline{2} = {-}{0.05em},
}
\textbf{Participant ID} & \textbf{Age} & \textbf{Gender} & \textbf{Onset} & \textbf{Level of Visual Impairment}        \\
P0 (Pilot study)                     & 67           & Female          & Acquired       & Legal blindness                            \\
P1                      & 24           & Female          & Acquired       & Legal blindness                            \\
P2                      & 30           & Female          & Congenital     & Total Blindness                            \\
P3                      & 23           & Female          & Congenital     & Total Blindness                            \\
P4                      & 28           & Female          & Congenital     & Blindness with some light/color perception \\
P5                      & 23           & Male            & Acquired       & Blindness with some light/color perception \\
P6                      & 44           & Female          & Congenital     & Legal blindness                            \\
P7                      & 24           & Male            & Acquired       & Blindness with some light/color perception \\
P8                      & 18           & Male            & Congenital     & Total Blindness                            \\
P9                     & 41           & Male            & Acquired       & Blindness with some light/color perception \\
P10                     & 23           & Male            & Congenital     & Total Blindness                            \\
P11                     & 32           & Male            & Congenital     & Total Blindness                            \\
P12                     & 33           & Female          & Congenital     & Total Blindness                            \\
P13                     & 38           & Female          & Congenital     & Total Blindness                            \\
P14                     & 33           & Male            & Acquired       & Legal blindness                            
\end{tblr}
\caption{Participant demographics for our user study}
\label{tab:participants}
\end{table*}}

\section{Sampled Videos}
\begin{table*}[!hb]
\begin{tabular}{lllll}
\hline
\textbf{Index} & \textbf{Topic} & \textbf{Duration} & \textbf{Resolution} & \textbf{Scenes}    \\ \hline
V1                              & Education                         & 0:01:53                            & 720p                                 & Grass, kitchen, restaurant                  \\
V2                              & Trailer                          & 0:02:10                            & 720p                                 & Library, playground                         \\
V3                              & People \& Blogs                      & 0:01:29                            & 720p                                 & Snow, street, bus                           \\
V4                              & Activism                      & 0:03:37                            & 720p                                 & Streets                                     \\
V5                              & Film \& Animation                         & 0:01:34                            & 720p                                 & Office                                      \\
V6                              & Comedy                            & 0:02:50                            & 720p                                 & Living room                                 \\
V7                              & People \& Blogs              & 0:02:08                            & 720p                                 & Workroom, streets                           \\
V8                              & Trailer    & 0:02:32                            & 720p                                 & Indoor rooms                                \\
V9                              & Education                       & 0:03:11                            & 720p                                 & Classrooms, library, playground             \\
V10                             & Entertainment                   & 0:04:58                            & 720p                                 & Sports field, factory field, indoor room    \\
V11                             & People \& Blogs                       & 0:01:55                            & 720p                                 & Living room, garden, corridor, sports field \\
V12                             & Film \& Animation              & 0:05:54                            & 720p                                 & Living room, street, bedroom                \\
V13                             & Pets \& Animals               & 0:02:27                            & 720p                                 & Living room, Kitchen, Bedroom               \\
V14                             & Activism                    & 0:02:39                            & 720p                                 & Living room                                 \\
V15                             & Comedy                          & 0:02:27                            & 720p                                 & Car, restaurant, home                       \\
V16                             & Trailor                          & 0:03:10                            & 720p                                 & Indoor rooms                                        \\
V17                             & Comedy                          & 0:01:49                            & 720p                                 & Living room, restaurant                     \\
V18                             & Sports                    & 0:02:41                            & 720p                                 & Ground, path, streets                       \\
V19                             & People \& Blogs                      & 0:05:00                            & 720p                                 & Riverside, laundry, classroom, home         \\
V20                             & Film \& Animation                        & 0:03:08                            & 720p                                 & Indoor rooms                                      \\ \hline
\end{tabular}
\caption{Videos used for technical evaluation. V1-V6 are used for our user study.}
\label{tab:videos}
\end{table*}

\twocolumn

\section{GPT-4 Prompt for Object Description Refinement}
\label{app:prompt}
We used the following prompt to refine each object-level description: \texttt{``Given the following visual-scene description [Frame-level description] and objects in the scene [A list of objects in the scene], refine the following object-level description to make sure: 1. The object is reasonable to exist in this scene. 2. The description style is natural and easy to understand. 3. Include relevant information from the visual-scene description.''}
\end{document}

%% file: 1-Introduction.tex
\section{Introduction}


Audio description (AD) offers spoken narration to visual elements in videos, films, or performances, making them accessible to individuals who are blind or low-vision (BLV). This adaptation is vital for BLV users to engage with video content as it deciphers visual content that may otherwise remain inaccessible~\cite{snyder2005audio,fryer2016introduction}. 

For professionally produced media content, such as movies and TV series, creating ADs requires significant collaborative efforts of a team of experts including content creators, audio description writers, narrators, and audio engineers. The landscape is different for user-generated content. In recent years, platforms such as YouTube, TikToks, and vlogs have witnessed exponential growth. However, according to the study in~\cite{Bartolome2023ALR}, the adoption rate of ADs on these platforms is significantly lower. In recent years, with advances in computer vision and natural language processing, the research community has introduced techniques for automatically generating ADs with end-to-end machine learning models~\cite{mehta2020automated,dingautoad}. However, their applicability remains limited due to accuracy concerns.



Even for well-crafted ADs in their existing format, several intrinsic limitations persist, due to their static nature.


\begin{enumerate}
    \item \textbf{Content Depth and Relevance:} Current AD norms underscore clarity and significance, leading to ADs primarily covering keyframes and pivotal video information~\cite{fryer2016introduction,rai2010comparative,vercauteren2007towards}. Meanwhile, as highlighted in~\cite{treemicrosoft}, the depth and quantity of content desired in an AD for a specific video frame can differ widely among users. Our study reaffirms this, indicating that users tend to explore different frames and objects in the same video. A gap thus emerges in conventional ADs between \textit{what a BLV user wants} and \textit{what ADs provide}, causing information loss and diminishing user engagement. This issue also affects fairness in information access, as information about some ``less important'' (as determined by the AD creators) objects is completely omitted from ADs, denying many BLV users the chance to perceive and acknowledge their presence.

    \item \textbf{Mental Load and Autonomy: } Prior research indicates that BLV audiences experience a higher mental load while consuming video content through ADs~\cite{fresno2014picture,holsanova2022cognitive} compared with sighted viewers. Specifically, BLV audiences have less autonomy when consuming video content, while sighted viewers are able to choose visual cues to focus on. They have to consume the original soundtrack from the video and the ADs concurrently through the audio modality. We argue that instead of allowing them to only passively \textit{hear} video content through \textit{static} ADs, an accessible video system should offer users more flexibility to interactively explore the details of video content at their own pace to reduce mental load. 

    \item \textbf{Immersive Experience: } Existing ADs focus primarily on providing information to facilitate user understanding of the content~\cite{fryer2016introduction,rai2010comparative,vercauteren2007towards}, while failing to offer immersive experiences to BLV audiences. In our study with 14 BLV participants, several reported that although ADs can facilitate the \textit{understanding} of the content, prolonged listening led to boredom, distraction, and disengagement.
\end{enumerate}

To address these issues, we present \textsc{Spica}\footnote{\textsc{Spica} is the acronym for \textbf{S}ystem for \textbf{P}roviding \textbf{I}nteractive \textbf{C}ontent for \textbf{A}ccessibility}, an AI-powered system that enables users to interactively explore video contents through layered ADs and spatial sound effects for individual objects in the frame. Instead of passively consuming predetermined frame-level ADs, users can actively explore individual objects within frames that they are interested in. They can access detailed object-specific ADs or sound effects through either a touch interface or a screen reader. Additionally, users can also perceive the location of objects through immersive spatial sound or descriptions of the object location within ADs. For users retaining some light or color perception, \textsc{Spica} further aids by overlaying a contrasting color mask on objects, enhancing visibility.

A unique design consideration of \textsc{Spica} is that it keeps existing ADs in video while augmenting the user's experience by providing users the agency to further explore video content based on their individual preferences. The extra information is automatically generated by a machine learning (ML) pipeline that detects and generates descriptions for keyframes for user exploration, identifies objects of interest in those frames, and generates descriptions or retrieves sound effects for these objects. The pipeline combines state-of-the-art computer vision techniques with large language models (LLMs) to ensure information consistency, improve the comprehensiveness of generated ADs, and elevate immersion in generated descriptions. The performance of the machine learning pipeline is evaluated through an offline benchmark evaluation.

To evaluate the usability and usefulness of \textsc{Spica}, we conducted a study with 14 users from the BLV community. These participants represented a wide range of BLV conditions. Study participants found \textsc{Spica} easy and natural to use, helpful in fulfilling their personal information access needs that were not met by conventional ADs, and useful in improving the immersive media consumption experience for BLV users. 

The development of \textsc{Spica} allowed us to use it as a tool to further understand the video information access needs and behaviors of BLV users. Specifically, we reported findings on the following aspects:
\begin{enumerate}
    \item How does the additional interactivity impact the way BLV users consumes videos?
    \item What are the mental processes when users engage with interactive strategies and layered information? Specifically, we investigated how their mental states change from a regular consumption experience to an interruption experience.
    \item What are the cues and approaches that users use to decide whether to pause videos for deeper exploration and with which objects they choose to engage?
\end{enumerate}
Two researchers of this paper analyzed the interview transcript through affinity diagramming and coded it with video recordings of the behaviors of the participants using established open-coding methods~\cite{brod2009qualitative}.
Drawing from the insights, we summarized the findings and design implications and identified several research opportunities for future work on advancing video interactivity for the BLV audience.

To sum up, we present three main contributions in this paper:
\begin{itemize}
    \item \textsc{Spica}, an AI-enabled interactive system that improves the understanding and the immersive viewing experiences of video content for BLV users.
    \item A user study with 14 BLV participants, validating the effectiveness and applicability of \textsc{Spica} in enhancing video consumption for BLV users.
    \item Insights and findings for future work on augmenting videos with interactivity for accessibility.
\end{itemize}

%% file: 2-Related_work.tex
\section{Related Work}
\label{sec:relatedwork}
  
\subsection{Audio Descriptions Generation}
\label{sec:relatedwork_ad_gen}

Audio descriptions are essential to foster an accessible user experience for individuals with visual impairments~\cite{snyder2005audio,fryer2016introduction} to consume video content. This ensures that visually impaired people can access, comprehend and engage with both new and existing video content. In an age where video serves as a primary medium for the dissemination of information across the news, education, and entertainment sectors, video accessibility has been an important part of social equity and technological inclusivity~\cite{packer2015overview}. Existing guidelines for audio descriptions advocate for precise, concise, and objective narrations that illuminate crucial visuals without interrupting or preempting the video dialogue or forthcoming content~\cite{youdescribeMiele,remael2015pictures,caldwell2008web}. However, the limited availability and high cost of professional audio description services make it challenging to widely implement such guidelines, thus constraining the accessibility of video content~\cite{thompson2017my}.

Previous research in the field of audio description (AD) generation has explored various approaches. Manual methods, for instance, include specialized authoring tools~\cite{branje2012livedescribe}, peer feedback mechanisms on audio descriptions~\cite{natalie2020viscene,saray2011adaptive}, accessibility verification checks~\cite{liu2022crossa11y}, and dedicated platforms for hosting these descriptions~\cite{youdescribeMiele}. Platforms like LiveDescribe \cite{branje2012livedescribe} and Rescribe~\cite{pavel2020rescribe} aim to democratize the generation of audio descriptions by offering user-friendly interfaces that even novices can leverage to produce audio descriptions with fewer resources~\cite{kobayashi2009providing}. Nevertheless, these manual methods can be inefficient for content creators and do not scale well to the enormous corpus of video content currently available.

Recent advances in deep learning and computer vision technologies have paved the way for automated approaches in AD generation~\cite{aafaq2019video,salisbury2017toward}. Previous studies have employed AI to accelerate AD creation, although with varying degrees of success~\cite{campos2020cinead,gagnon2010computer,gagnon2009towards, wang2021toward, yuksel2020human}. However, these automated methods can sometimes be imprecise and lack the contextual richness needed for an audience with visual impairments. This issue mirrors the limitations observed in automatically generated alternative texts for images~\cite{wu2017automatic}. In contrast to existing approaches, our system aims to automatically generate hierarchical ADs that offer more granular details. Furthermore, our system empowers users with visual impairments by granting them greater control over the viewing experience, including the ability to request additional descriptive details as needed.

\subsection{Interactive Exploration for Visual Content}

Interactive visual content consumption approaches have been widely used in domains such as education, media entertainment, healthcare, etc. to improve users' understanding~\cite{Ijaz2020PlayerEO,Sharma2014MultiuserVC,Xu2023HeritageSiteAA}. In tools for BLV users, exploration-based methods are also widely applied. 

Previous work such as ImageExplorer~\cite{lee2022imageexplorer} focused on image exploration: a hierarchical strategy is designed to allow BLV users to enhance user understanding of a static picture. Similarly, Slidecho~\cite{peng2021slidecho} extracted text and image elements from the presentation and aligned these elements to the presenter's speech to provide BLV learners with more organized information. Unlike previous work, \textsc{Spica} targets video consumption, which introduces unique challenges. First, the narration of a video usually follows a specific time order; therefore, a tool should be able to support the users' comprehension of the scene transitions while consistently interpreting non-visual elements such as objects and frame details to the user. Additionally, unlike static images and slide exploration which concentrate on improving the \textit{understanding}, video content exploration also needs to optimize for users' engagement~\cite{richardson2020engagement,fryerPresense}. Another relevant work is AVscript~\cite{avscript}, which provides support for non-visual video \textit{editing}, which features different challenges, goals, and expected expertise of target user groups.



Other works also focus on enabling BLV participants to explore real-world objects. For instance, Seeing AI\footnote{\url{https://www.microsoft.com/en-us/ai/seeing-ai}} helps BLV individuals gain information about their surroundings, including visual scenes, documents, and objects through the camera of their mobile devices. CueSee~\cite{cuesee} utilizes visual cues to guide low-vision users to explore different products using a Head-Mounted Device. Earlier work such as Slide Rule~\cite{kane2008slide} designs a set of audio-based multi-touch interaction techniques that enable blind users to access touchscreen applications. RegionSpeak~\cite{zhong2015regionspeak} applies a touch-based method to enable BLV users to explore the image by regions annotated by Amazon Mechanical Turk workers to address information loss in image exploration. 

Different from prior work, our system enables users to explore the visual scene temporarily and also allows users to delve into the frame to explore different visual objects in videos. Through our study, we identified special challenges and summarized design insights when applying an interaction-based method to video content.

\subsection{BLV User Engagement with Visual Content}

User engagement while using the tool is as important as how they use the tool to understand the visual content itself. Prior research such as~\cite{liu2021makes,gleason2020making} has shown that BLV users encounter various problems that prevent them from immersion and enjoyment while accessing the visual content. For instance, they would feel mentally tired after using the ADs for a period of time. It is also shown that low-quality ADs would result in users' confusion and tiredness. Furthermore, the research in~\cite{kobayashi2010synthesized} found that, although ADs give users more information and improve their understanding, they fall short in increasing the spatial awareness of BLV users, thus decreasing their overall engagement. There are other factors that would also affect BLV users' engagement while experiencing visual content, for example, details in the visual scene~\cite{matamala2015audio}, the delivery style of the AD~\cite{matamala2007designing,holsanova2016cognitive}, and emotion impact~\cite{caro2016testing}.

In \textsc{Spica}, the system provides users with object details in the visual scene to help BLV users engage in the video content. To increase engagement while exploring, inspired by video dubbing technology, we associated a sound effect with the object while the object-level description was playing. To further strengthen the spatial awareness of the BLV users while also decreasing their cognitive load, we encode the 3D position of an object into a sound effect. 

%% file: 3-System.tex
\section{SPICA System}
\label{sec:sys}

\begin{figure*}[htb]
  \includegraphics[width=0.9\linewidth]{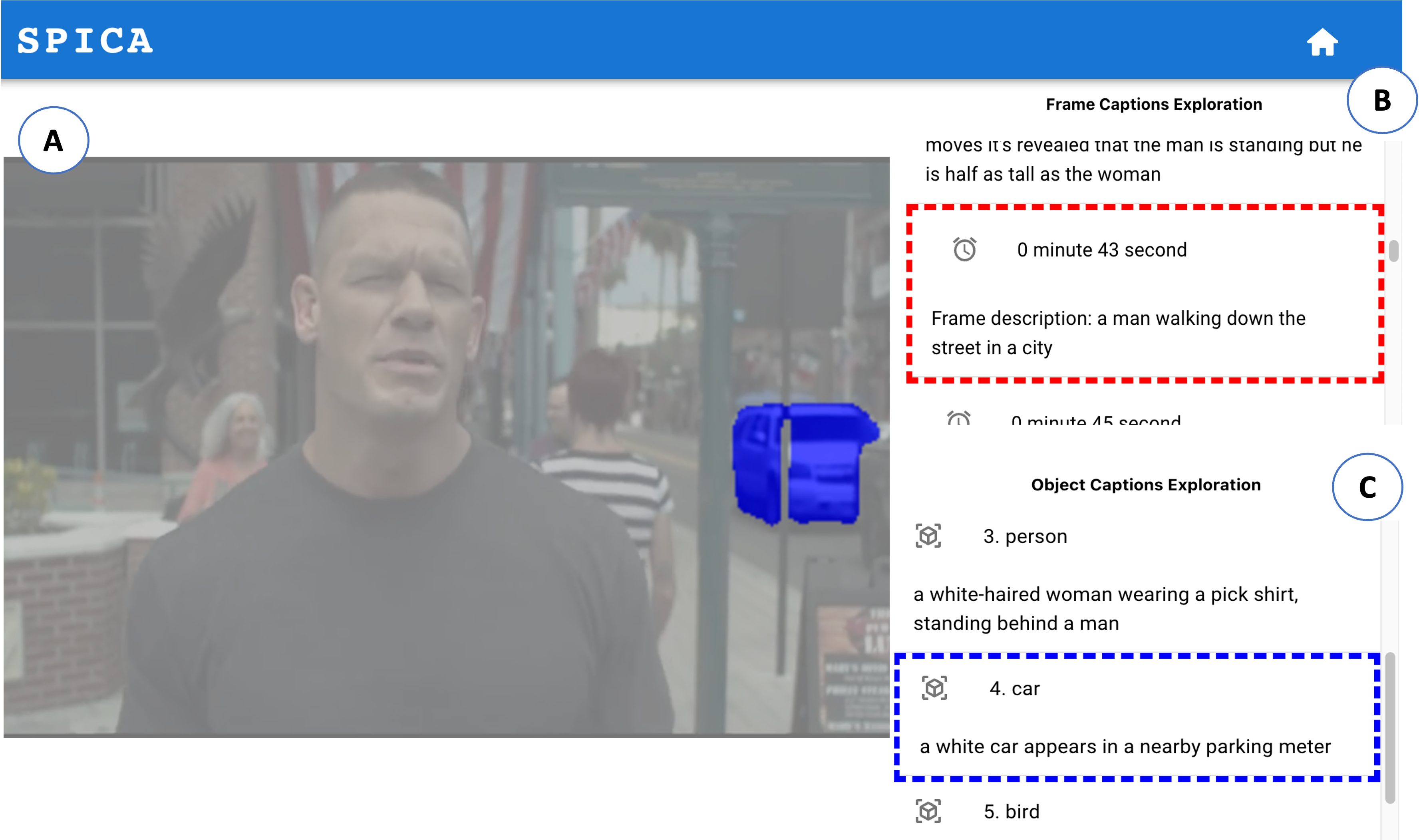}
  \caption{The main interface of \textsc{Spica}. \textsc{Spica} provides interactivity for BLV users to explore the video. A) The video player. Users can explore objects in the frame using fingers or keyboard arrow keys. The object would be highlighted with a high-contrast color mask if selected. B) Frame-Level Caption List. Users can use arrow keys on the keyboard to navigate to different visual scenes. C) Object-Level Caption List. Users can use arrow keys to scan through objects. Once the object is selected, a spatial sound effect associated with the object will be played based on its estimated 3D position.}
  \label{fig:system_ui}
  \Description[The main interface of SPICA]{In the main interface of SPICA, there is a video pane taking up more than half of the interface to the left. On the right side, anther pane shows the frame-level descriptions for the identified frames with timestamps. Below there is another pane displays object-description for the current frame.}
\end{figure*}

We introduce \textsc{Spica}, a system that enables BLV users to interactively explore video content through augmented ADs. This section starts with an overview that demonstrates an example workflow for a user with \textsc{Spica}. Then, we give an overview of \textsc{Spica}'s system architecture, and describe three key interactive features that enable BLV users to (1) temporally explore key frames of interest on the timeline and (2) spatially explore objects of interest within each frame. This section ends with an explanation of \textsc{Spica}'s implementation details.


\subsection{Example Workflow}
\label{sec:usage_scene}

To understand how \textsc{Spica} operates, let us consider an example involving Lucy. Lucy has experienced a significant loss of vision, retaining only minimal light and color perception. Apart from this, she has no other vision impairments.

Lucy wants to watch a baking tutorial video using \textsc{Spica}. She started the video by pressing the \texttt{SPACE} key on her keyboard. As the video progressed to a segment with an extended native AD (i.e., the original descriptive narration provided by the video maker or AD creator), \textsc{Spica} automatically paused the video and vocalized the AD: \textit{``A young man is standing behind the table holding a pot.''}

If Lucy chooses not to use any interactive features of \textsc{Spica}, the video will play exactly the same way as normal videos with native ADs. However, when she heard that AD, she was intrigued to learn more about how the young man and the pot look like. Moreover, she was curious to learn if there are other objects on the table. Therefore, she decides to further explore this frame. To do this, she used her finger to scan over the video frame. She first touched the pot, simultaneously, \textsc{Spica} automatically provided an audio caption: \textit{``This is a pot, located at the top right, description: a silver metal pot with melted butter inside.''} Meanwhile, a colored overlay highlighted the pot's shape and location in the video frame. This visualization aids Lucy in discerning the pot's shape, leveraging her residual color perception. Additionally, a spatial audio effect mimicking boiling is playing in spatial audio to further enhance her spatial perception and immersive sensory experience.

Lucy figured that she understood the current visual scene better; however, she wanted to ensure that she did not miss any important cooking steps until then. To do this, she pressed the \faIcon{arrow-circle-up} key on her keyboard and \textsc{Spica} automatically took her to the previous keyframe that the system identified and read the description of that frame at the same time. She can also press the \faIcon{arrow-circle-down} key to go back to the frame where she just paused, or keep pressing the arrow key to forward to other frames. 

Then, Lucy can choose to explore this frame in detail or continue watching. In addition to exploring by touch on the screen, should she wish to engage in a hands-free exploration experience, Lucy can employ screen reader controls to navigate through objects in a frame. To do this, she hits the \texttt{RETURN} key to switch to this mode (she can simply press \texttt{RETURN} again to switch back). In this mode, using the up \faIcon{arrow-circle-up} and down \faIcon{arrow-circle-down} arrow keys allows her to cycle through objects present in the frame. Same as the touch exploration mode, for any selected object, \textsc{Spica} plays its object-level description, offers a relevant spatial sound effect, and displays a contrasting color overlay. To proceed, Lucy simply presses \texttt{SPACE}.

\subsection{System Overview} 
\label{sec:arch}



\begin{figure*}
    \centering
    \includegraphics[width=0.9\textwidth]{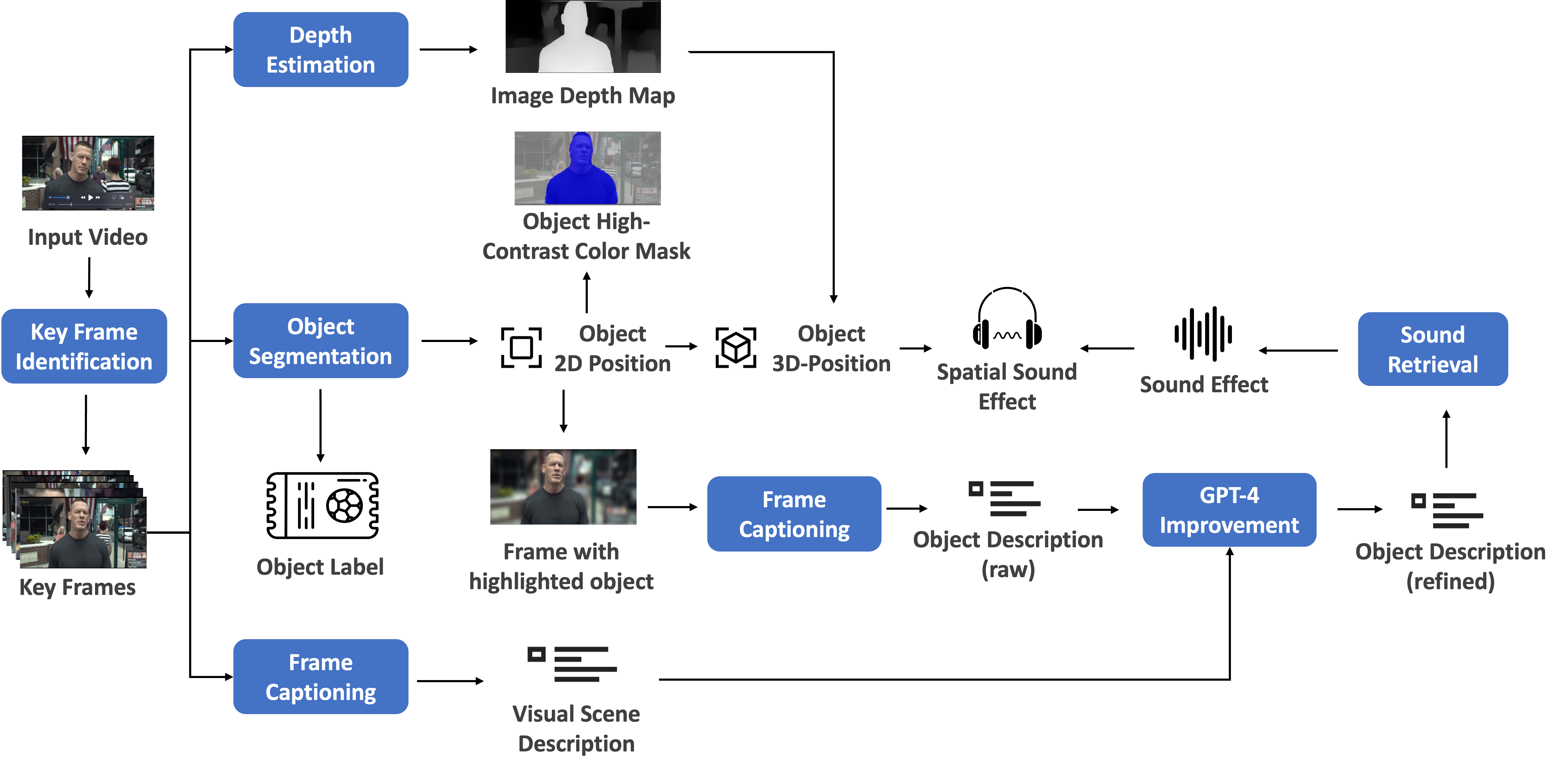}
    \caption{The machine learning pipeline used in \textsc{Spica}.}
    \label{fig:ml_pipl}
    \Description[The machine learning pipeline used in {Spica}.]{A flowchart shows how the backend algorithm of SPICA analyzes the video and obtain the estimated object 3D positions, frame-level descriptions, object-level descriptions, and spatial sound effects.}
\end{figure*}
\textsc{Spica} consists of the following two blocks:
\begin{enumerate}[leftmargin=0pt]
\setlength\itemsep{0.5em}
    \item[] \textbf{ML Pipeline:} The ML pipeline (Fig.~\ref{fig:ml_pipl}) consists of several modules for producing the layered frame-level descriptions, object-level descriptions, high-contrast color mask, and spatial sound effects: (1) \textit{Scene analysis}: This module identifies key frames in the video and generates/retrieves the corresponding visual scene descriptions for each key frame; (2) \textit{Object segmentation}: Within each frame, it detects and segments objects of interests; (3) \textit{Object description generation}: Provided the object 2D position, it first processes the corresponding frame with the object highlighted, and generates the raw object description. Then it generates a detailed natural language description for the object that is consistent with other objects in the context of the scene through GPT-4 using the scene description and the raw description; (4) \textit{Object sound effect retrieval}: Given the object description, it retrieves the most appropriate sound effect from \textit{freesound}~\footnote{\url{https://freesound.org}}, a large collaborative database of licensed sounds, that represents the object; and (5) \textit{Object depth estimation}: For each identified object, the model predicts its depth position relative to the frame. The depth position, together with the predicted 2D position are used to enhance the generated object description and the retrieved sound effect with spatial effects.

    \item[] \textbf{Frontend Interactive Application: } There are three main components in the frontend UI: (1) \textit{Video player} (Fig.~\ref{fig:system_ui}-A):  It enables video playback and allows users to explore objects within frames using touch gestures or arrow keys on the keyboard. The player also renders a high-contrast color overlay when an object is selected; (2) \textit{Frame-level caption list} (Fig.~\ref{fig:system_ui}-B): It presents the timestamps of the identified keyframes, alongside their respective scene descriptions. Users can navigate through these frames using the arrow keys; (3) \textit{Object-level caption list} (Fig.~\ref{fig:system_ui}-C): It displays labels and descriptions of individual objects within the selected frame, allowing users to explore each object in detail.
    
\end{enumerate}


\subsection{Temporal Navigation on Visual Scenes}
\label{sec:temp_exp}


We designed a temporal navigation mechanism to support users in transitioning between distinct visual scenes within a video. \looseness=-1


Traditional video players typically offer users the basic functionality to skip backward or forward by a predetermined time span. \textsc{Spica} goes further by enabling BLV users to jump directly to specific scenes. This design decision was informed by the insights of the study in~\cite{avscript}, which found that BLV users often find it more intuitive to navigate videos based on scene descriptions rather than timelines.

Drawing inspiration from this, \textsc{Spica} allows users to use arrow keys to shift to either the preceding or subsequent keyframe that features a unique visual scene. As they do, the video automatically adjusts to the relevant timestamp. To ensure that users remain oriented throughout their viewing experience, \textsc{Spica} announces the current timestamp and reads out the scene-level description each time a transition to a new frame occurs. This feature provides users with a continuous sense of their progression within the video. The mechanism is designed for providing BLV users with greater flexibility and fostering a more detailed and comprehensive understanding of the video narrative.




\subsection{Spatial Exploration of Objects within Frames}
\label{sec:spatial_exp}
Previous studies~\cite{lee2022imageexplorer,morris2018rich} have shown that enabling BLV users to actively interact with visual content (static images in these studies) in a hierarchical manner can improve their comprehension while reducing information overload. In line with this finding, \textsc{Spica} incorporates strategies that facilitate users exploring and examining objects within a video frame spatially.
Our system introduces two primary exploration techniques:

\begin{enumerate}
    \item \textbf{Touch-Based Exploration:} Users can interact directly with the video frame. This not only provides the 2D position of the object but also offers an intuitive feel for the object's size.
    \item \textbf{Keyboard-Based Exploration:} By using arrow keys, users can sequentially sift through objects within the frame. This method ensures that users can comprehensively identify every element in the visual scene and easily reposition to a particular object.
\end{enumerate}

Users can alternate between these two techniques as needed. Upon selecting an object, its description is both audibly read out and visually presented in a dedicated description panel. To streamline the exploration process and minimize redundancy, especially when multiple objects are similar to each other, each description starts with a unique index number, ranging from 1 to n (where `n' denotes the total objects within the frame). Furthermore, to aid low-vision users who retain some light or color perception, a vivid, contrasting mask highlights the selected object in the frame. 


Informed by recent research~\cite{frontrow,gaurav2023front} that adopted spatial audio as a clue to enhance the viewing experience of BLV users on sports broadcasts, we designed an object-specific spatial audio technique to enhance user spatial awareness and increase engagement. The sound effect is retrieved using the keywords in the object description generated by the ML pipeline (details in Section~\ref{sec:caption_gen_pipl}) through the freesound API~\footnote{\url{https://freesound.org/docs/api/}}. The retrieved sound effect will then be spatialized based on the object's predicted 3D position. When an object is selected, \textsc{Spica} plays the associated spatial sound effect.


\subsection{Keyframe Detection and Description Generation Pipeline}
\label{sec:caption_gen_pipl}
We constructed a machine learning pipeline (Figure~\ref{fig:ml_pipl}) analyze the video in both frame- and object-level, and identify keyframes. Noticeably, although models like~\cite{islam2023efficient,wu2022scene} can detect scene transitions in an end-to-end way, their priority is based on the visual similarity of different scenes. On the contrary, our aim is to offer users a richer exploration of varied objects, ensuring that keyframes are densely populated throughout the video for comprehensive exploration. As such, our pipeline adopts the following rules to segment the video:

\begin{enumerate}
\item \textbf{Native audio descriptions (AD) exists:} The video is segmented at frames containing original audio descriptions that were built into the original video.
\item \textbf{Visual scene varies:} A segment is made if the visual scene description of a frame differs significantly from the previous frame.
\item \textbf{Object information changes:} Segmentation occurs if the present frame and its predecessor differ significantly in object count and types.
\item \textbf{Reaches maximum slice interval:} We set a cap on the interval duration at 5 seconds. If this threshold is reached without a new cut, the video is automatically segmented.
\end{enumerate}

After segmenting the video into chunks of frames, the pipeline identifies a key frame for each chunk based on the information a frame contains. The informative level is measured by considering a frame's \textit{object score} and \textit{semantic score}. Specifically, the \textit{object score} is computed by normalizing the object count in the frame with the minimum and maximum object counts within the group. It measures the object diversity of a frame. The \textit{semantic score} evaluates the uniqueness of a frame's visual scene compared to others in its group. This is done by analyzing the textual similarity of each frame's visual description. We used Sentence-BERT~\cite{reimers2019sentence} to compute such similarity. Noticeably, if a frame that has a native AD is included in a chunk, the frame is considered a key frame by default.


We utilize an off-the-shelf image captioning model~\cite{wang2022ofa} to provide descriptions for key frames lacking the original AD. For individual object descriptions, this model is repurposed: instead of generating description for the cropped object only, it generates captions using a version of the frame where the background of the object is blurred, thereby emphasizing context from the main visual scene. Afterward, we refine the object-level captions with GPT-4 by prompting it to improve the consistency between the initial object captions and frame captions to make them more accessible to the visually impaired. Three illustrative examples of this refinement can be found in Table~\ref{tab:qualitative_eval}. The prompt used for this transformation can be found in Appendix~\ref{app:prompt}.

\subsection{Implementation}
\label{sec:sys_implement}
We developed the web-based interactive system of \textsc{Spica} using React\footnote{\url{https://react.dev}} for the front-end, and Flask\footnote{\url{https://flask.palletsprojects.com}} for the back-end. We used Google text-to-speech\footnote{\url{https://cloud.google.com/text-to-speech}} API to read out the generated descriptions.

In the machine learning pipeline (detailed in Section~\ref{sec:caption_gen_pipl}), the object detection used the Mask RCNN model~\cite{he2017mask}. The object detector identifies only significant objects in the video frame, using a pre-trained model on the MS COCO dataset~\cite{lin2014microsoft} that contains 80 common object categories e.g., people, car, bird. Furthermore, we apply a manual threshold of 0.9 for the confidence score to filter out predictions with low confidence. Additionally, we filter out small objects whose height and weight are less than 0.1 percent of the height or width of the frame. The object descriptions displayed in the interface are ordered by the 2D area of the corresponding object's mask, relative to the size of the visual scene.

In addition, the image captioning model is adapted from~\cite{wang2022ofa}, and the depth estimation model is based on~\cite{kim2022global}. Notably, these models are effective for our use case without necessitating additional fine-tuning.


%% file: 4-ML_pipeline.tex
\section{Technical Evaluation}
\label{sec:tech_eval}
To validate the technical feasibility of the pipeline described in Section~\ref{sec:caption_gen_pipl}, we conducted an offline benchmark evaluation. We evaluated the accuracy of object label predictions and the quality of object-level descriptions generated by the pipeline.

For comparison, we used a baseline pipeline that directly connects a state-of-the-art object detection model~\cite{he2017mask} to a state-of-the-art image captioning model~\cite{wang2022ofa}. Both models used were one of the most robust and widely used open-source methods for their tasks as of the year 2023. 

\subsection{Dataset}
The evaluation was conducted on a sample subset of the QueryYD dataset\footnote{\url{https://www.robots.ox.ac.uk/~vgg/data/queryd/}} with 20 videos. Detailed information about each video is shown in Table~\ref{tab:videos}. These videos were originally sourced from YouTube and came with audio descriptions contributed by sighted volunteers from the YouDescribe project\footnote{\url{https://youdescribe.org}}. The sample set of videos are within six minutes and in English; each video has a minimum resolution of 720p and has complete native audio descriptions. The sampled videos cover a diverse set of topics, including movie clips, vlogs, clips of TV shows, advertisements, etc.
 
\subsection{Procedure}

We began by acquiring a set of video frames for each video. We used the keyframe detection mechanism in our pipeline (Section~\ref{sec:caption_gen_pipl}) to obtain frames with distinct visual scenes in the video. Then, for each frame, we use both our approach and the baseline approach to get the object labels and descriptions for the objects. 

To evaluate the precision and recall of object labels, two annotators manually went through each frame to annotate the label and position of each object to obtain the ground-truth results using LabelMe\footnote{\url{http://labelme.csail.mit.edu/Release3.0/}}. \textit{Precision} is computed by dividing the number of correctly labeled objects by the total number of objects the model identifies. It reflects the proportion of objects accurately labeled by the model. \textit{Recall} is formulated as the fraction of the intersection of objects accurately labeled by the model and those labeled in the annotated ground truth, divided by the total number of objects labeled in the annotated ground truth. It measures the overlap between objects the model correctly identifies and those present in the annotated ground truth. As our pipeline does not optimize the label prediction process compared to the baseline, we did not report a comparison.

To evaluate the quality of the object-level descriptions, two graders manually rated each description on a seven-point scale, where ``1'' represents the worst quality and ``7'' means the best quality. They did not know the pipeline that generated each description. The final score for each description is the average score of two graders. The graders rated each caption considering its correctness, naturalness, and comprehensiveness. There was a strong inter-grader agreement with a Cohen's weighted kappa 0.71.

\subsection{Results}
\subsubsection{Labels precision and recall} The overall precision of our pipeline is 0.939, and the SD across all videos is 0.03. The recall of our pipeline is 0.791, with an SD of 0.07 across all videos. The incorrect and missing predictions mainly came from occlusion in the frame, or because the object fell into a category that was missing in the training set of the model. Furthermore, each video exhibited a similar level of precision and recall, which further shows the robustness of our pipeline in different video scenarios. The result is shown in Table~\ref{tab:acc_recall}.

\begin{table}[htb]
\begin{tabular}{ccc}
\hline
         & \textbf{Mean}  & \textbf{SD} \\ \hline
\textbf{Precision} & 0.939 & 0.03     \\
\textbf{Recall}   & 0.791 & 0.07     \\ \hline
\end{tabular}
\caption{Label precision and recall of our pipeline.}
\label{tab:acc_recall}
\end{table}


\subsubsection{Descriptions quality} The baseline condition achieved an average rating of 3.91, with a SD of 1.24. In contrast, our approach achieved a higher average rating of 5.10, with a SD of 1.29. A one-way ANOVA test ($\alpha$=0.05) showed a significant difference between the average ratings of the two approaches. The result indicates that our pipeline can deliver higher-quality descriptions to human users. The result is shown in Table~\ref{tab:anova_pipl}

\begin{table}[htb]
\begin{tabular}{ccc}
\hline
                 & \textbf{Average Rating}   & \textbf{SD} \\ \hline
\textbf{Baseline}    & 3.91             & 1.24     \\
\textbf{Ours}             & 5.1***              & 1.29  \\
\hline
\end{tabular}
\caption{Average quality ratings for the baseline pipeline and our pipeline (***p<0.001).}
\label{tab:anova_pipl}
\end{table}

Table~\ref{tab:qualitative_eval} shows three examples to illustrate the change in object-level descriptions using our pipeline. Results produced by our pipeline included more objects with individual descriptions compared with the original frame description. Meanwhile, compared to the baseline pipeline, ours delivers more natural and more consistent descriptions by removing irrelevant phrases and aligning the object-level description with the frame-level description. For instance, in example 1, our pipeline enhances the baseline description by replacing the subject's identifier in the baseline description (``woman'') with a concrete name (``Rachel'') inferred from the context and adding position information (``behind Ross'') obtained from the original frame caption.

\aptLtoX[graphic=no,type=html]{\begin{table}
\centering
\begin{tabular}{p{6pc}|c|p{13pc}|p{13pc}}
\toprule
\textbf{Frame Caption}                                & \textbf{Object Label} & \textbf{Object Description (baseline)}                          & \textbf{Object Description (ours)}                                                          \\
\midrule
Rachel appears behind Ross in his doorway.            & person                & A woman is wearing a black and white dress, she has blond hair. & \textcolor{blue}{\textbf{Rachel}}, a woman with blonde hair, \textcolor{blue}{\textbf{is appearing behind Ross}} in a black and white dress. \\
                                                      & person                & A man is standing in a room wearing a black suit.                & \textcolor{blue}{\textbf{Ross}}, a man in a black suit, \textcolor{blue}{\textbf{is standing in a doorway in a dark room}}.                   \\
A man in a black shirt is walking down a city street. & person                & A man wearing a black shirt walking down the street.             & A man in a black shirt strolling along the city street.                                 \\
                                                      & person                & A man sitting by the street holding a bottle in his hand         & Another man is sitting by the street, holding a bottle in his hand.                     \\
                                                      & \textcolor{orange}{\textbf{stop sign}}             & A red stop sign is covered by some snow~                         & \textcolor{blue}{\textbf{In front of the man}}, there is a red stop sign, covered by some snow.~                   \\
                                                      & \textcolor{orange}{\textbf{car}}                   & \textcolor{red}{\textbf{A close up of}} a car is parking by the street.                    & A white sedan is parked by the street.                                                 \\
                                                      & \textcolor{orange}{\textbf{bus}}                   & \textcolor{red}{\textbf{A blurry image of}} a bus driving down the snowy street.~          & A bus is driving on the city street.                                                    \\
The scene changes to the man's living room            & person                & A person wearing a shirt and carrying a large black suitcase     & A man is dressed in a casual outfit, carrying a suitcase.                               \\
                                                      & \textcolor{orange}{\textbf{suitcase}}              & \textcolor{red}{\textbf{A blurry photo of}} a large black suitcase                         & A large black suitcase carried by the man.                                              \\
                                                      & \textcolor{orange}{\textbf{chair}}                 & A wooden chair, {possibly in an office}.                            & There is a wooden chair {in the living room}.                                            \\
                                                      & {potted plant}          & A plant in a pot on a window sill                                & A plant in a pot on a window sill                                                       \\
                                                      & {potted plant}          & A blue vase with a green plant in it                             & Another green plant in a blue pot \\ \hline                                                       
\end{tabular}
\caption{Object-level descriptions. The difference in descriptions generated by our pipeline, the baseline pipeline, and the objects that are missing in the frame description are highlighted. \textcolor{orange}{Orange: Missing objects in the frame caption;} \textcolor{red}{Red: Inaccuracies from the baseline pipeline that our pipeline fixed;} \textcolor{blue}{Blue: Additional details produced by our pipeline.}}
\label{tab:qualitative_eval}
\end{table}}
{\begin{table*}
\centering
\scalebox{0.9}{
\begin{tblr}{
width = \textwidth,
  row{1} = {c, font=\bfseries},
  colspec = {X[c,0.2\textwidth] X[c,0.15\textwidth] X[c,0.3\textwidth] X[c,0.3\textwidth]},
  cell{2}{1} = {r=2}{},
  cell{4}{1} = {r=5}{},
  cell{9}{1} = {r=5}{},
  vline{2,3,4} = {solid},
  hline{1,14} = {-}{0.08em},
  hline{2,4,9} = {-}{},
}
\textbf{Frame Caption}                                & \textbf{Object Label} & \textbf{Object Description (baseline)}                          & \textbf{Object Description (ours)}                                                          \\
Rachel appears behind Ross in his doorway.            & person                & A woman is wearing a black and white dress, she has blond hair. & \textcolor{blue}{\textbf{Rachel}}, a woman with blonde hair, \textcolor{blue}{\textbf{is appearing behind Ross}} in a black and white dress. \\
                                                      & person                & A man is standing in a room wearing a black suit.                & \textcolor{blue}{\textbf{Ross}}, a man in a black suit, \textcolor{blue}{\textbf{is standing in a doorway in a dark room}}.                   \\
A man in a black shirt is walking down a city street. & person                & A man wearing a black shirt walking down the street.             & A man in a black shirt strolling along the city street.                                 \\
                                                      & person                & A man sitting by the street holding a bottle in his hand         & Another man is sitting by the street, holding a bottle in his hand.                     \\
                                                      & \textcolor{orange}{\textbf{stop sign}}             & A red stop sign is covered by some snow~                         & \textcolor{blue}{\textbf{In front of the man}}, there is a red stop sign, covered by some snow.~                   \\
                                                      & \textcolor{orange}{\textbf{car}}                   & \textcolor{red}{\textbf{A close up of}} a car is parking by the street.                    & A white sedan is parked by the street.                                                 \\
                                                      & \textcolor{orange}{\textbf{bus}}                   & \textcolor{red}{\textbf{A blurry image of}} a bus driving down the snowy street.~          & A bus is driving on the city street.                                                    \\
The scene changes to the man's living room            & person                & A person wearing a shirt and carrying a large black suitcase     & A man is dressed in a casual outfit, carrying a suitcase.                               \\
                                                      & \textcolor{orange}{\textbf{suitcase}}              & \textcolor{red}{\textbf{A blurry photo of}} a large black suitcase                         & A large black suitcase carried by the man.                                              \\
                                                      & \textcolor{orange}{\textbf{chair}}                 & A wooden chair, {possibly in an office}.                            & There is a wooden chair {in the living room}.                                            \\
                                                      & {potted plant}          & A plant in a pot on a window sill                                & A plant in a pot on a window sill                                                       \\
                                                      & {potted plant}          & A blue vase with a green plant in it                             & Another green plant in a blue pot                                                       
\end{tblr}}
\caption{Object-level descriptions. The difference in descriptions generated by our pipeline, the baseline pipeline, and the objects that are missing in the frame description are highlighted. \textcolor{orange}{Orange: Missing objects in the frame caption;} \textcolor{red}{Red: Inaccuracies from the baseline pipeline that our pipeline fixed;} \textcolor{blue}{Blue: Additional details produced by our pipeline.}}
\label{tab:qualitative_eval}
\end{table*}}

%% file: 5-User_study.tex
\section{User Study}
\label{sec:usr_evaluations}
We conducted a within-subjects lab study\footnote{The study protocol was reviewed and approved by the IRB at our institution.} with 14 BLV participants to consume video content using \textsc{Spica} and a baseline condition. Specifically, the study seeks to answer the following research questions.

\begin{itemize}
    \item[RQ1.] What is the user peception to the usability of \textsc{Spica}?
    \item[RQ2.] How do individual interactive features of \textsc{Spica} influence users' information access and sense of immersion when consuming video content?
    \item[RQ3.] How would BLV users use \textsc{Spica} to explore video content, especially to determine when to pause and what to explore?
\end{itemize}

\subsection{Participants}
We recruited 14 BLV participants (P1-P14 in Table~\ref{tab:participants}) from social media and university mailing lists for this study. Their demographic information is shown in Table~\ref{tab:participants}. The average age of the participant group is 32.1 (SD=12.2). Our sample group of participants had various vision impairments ranging from legal blindness, blindness with some light/color perception, and total blindness. Each participant was compensated \$60 USD for their time.

\subsection{Study Design}
The study was conducted remotely on Zoom. We ask users to use tablets with keyboards or laptops with touchscreen to access the system so that they can easily alter between the two object exploration approaches (touch-based exploration and keyboard-based exploration, as detailed in Section~\ref{sec:spatial_exp}).

\subsubsection{Conditions}
The study used a within-subjects design. Each participant consumed two different videos in two conditions. In the baseline condition, the participant consumed the video content through the original ADs. In the experimental condition, users consumed the video using \textsc{Spica}. The order in which participants experienced the two conditions was counterbalanced.

\subsubsection{Videos}
We selected six videos (V1-V6 in Table~\ref{tab:videos}) for the user study from the QueryYD dataset. Each video was inspected prior to the study to ensure it had a cohesive storyline and high-quality original AD descriptions. The videos we chose ranged from 1.5 to 3 minutes in length. Additionally, each video was prepared under both specified conditions.



\subsubsection{Study procedure} 
Before the study session, each participant was asked to complete a pre-questionnaire about their demographics, their usage of accessible tools such as VoiceOver, NVDA, and Jaws, and their previous experience of video content consumption.

Each study session lasted around 60 minutes. At the beginning of each study session, after the consent process, one researcher reviewed the participant's responses to the pre-study questionnaire and discussed with the participants about their past experiences of consuming video content through ADs.

Following the tutorial, the participant proceeds to the video consumption task. Each participant will be randomly assigned two videos, one for each condition. The order of the two conditions was counterbalanced. For each condition, the researcher first conducted a tutorial session that lasted around 15 minutes, where the researcher walked through the participant on how to use the system features. After each video, participants are asked to rate their understanding and perceived immersion level on a 7-point Likert scale. This process is repeated until each participant has viewed a total of four videos, two under each condition. The participants are asked to try to understand the video content in their most comfortable way. We did not set a time limit for the video consumption session. 


After consuming the videos, each participant completed a post-study questionnaire, where they evaluated various statements concerning \textsc{Spica}'s usability and usefulness on a 7-point Likert scale. Lastly, the researcher conducted a 15-minute semi-structured interview with the participants to gather in-depth feedback about their experiences with both conditions, providing a comprehensive understanding of their reflections and potential improvements for the system. 


\subsubsection{Qualitative analysis procedure}
\label{sec:qualitative_method}
{Two authors of this paper conducted a qualitative analysis of the transcript and video recordings to analyze the participant feedback and the behaviors of the participants.}

{For the transcript, the researchers utilized affinity diagramming to put the quotes of the participants onto sticky notes using an online whiteboard. The researchers then collaboratively grouped these notes based on the thematic similarities of the quotes.}

{To further understand the active behaviors of the user (pauses and explorations) (RQ3), two researchers independently coded the notable participant behaviors. The coding process was conducted based on the following hypotheses: 

\begin{itemize}
    \item [H1:] Users' active behaviors are correlated to specific playback positions in a video.
    \item [H2:] User-initiated explorations happen when users perceive missing or conflicting information in a video.
\end{itemize}

Based on these hypotheses, for each recorded study session, each researcher annotated the following properties of a user action: (1) The temporal distance (in seconds) between the action and the closest original AD. (2) Whether the action occurred in a scene that is relevant to the scene described by the closest original AD. This is subjectively determined by each annotator based on the position of the video and the original AD. (3) Whether the user explored the information of the objects inside the frame. (4) Which object exploration method(s) did the user use. (5) What new audiovisual information is present in this scene compared to adjacent scenes.

We used established open-coding methods~\cite{brod2009qualitative}. Non-agreement cases were discussed to reconcile differences for establishing a cohesive codebook.} Using the codebook, we conducted a thematic analysis. These themes were then consolidated and evolved into the study findings, which are detailed in Section~\ref{sec:results_findings}.


\begin{figure*}
    \centering
    \includegraphics[width=0.9\textwidth]{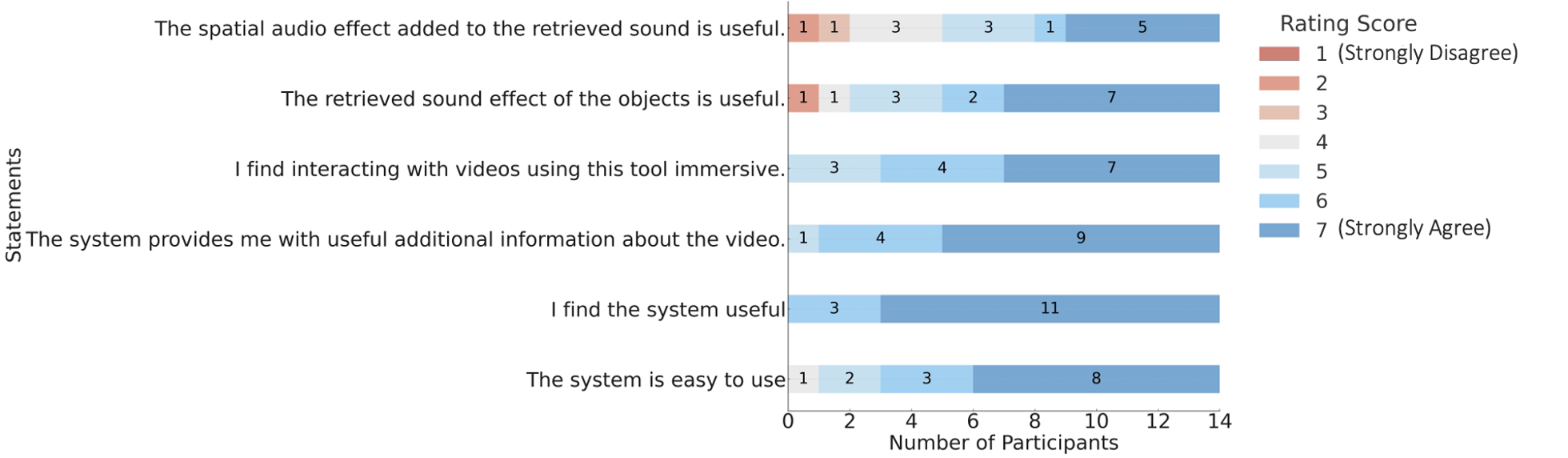}
    \caption{Participants' ratings to the usability and usefulness questions to \textsc{Spica}}
    \label{fig:post_quesionnaire}
    \Description[A diagram showing the summary of user ratings to the usability questions.]{The distribution of user ratings for each usability question in SPICA shows that the highest frequency is for the rating "7", indicating strong agree.}
\end{figure*}

\section{Results and findings}
\label{sec:results_findings}
\subsection{System Usability and Usefulness}
We assessed \textsc{Spica}'s usability and usefulness through the responses of the users in the post-study questionnaire (responses are shown in Fig.~\ref{fig:post_quesionnaire}) and their qualitative feedback in the interviews. 

\vspace{0.5em}
\noindent\textbf{{Quantitative evaluation.}} Generally, the participants found \textsc{Spica} effective and easy to use. Participants rated \textit{``The system is easy to use''} with a mean score of 6.29 (SD=0.99). The participants also found the system useful ($\mu=6.79$, SD=0.43) and provided them with additional information ($\mu=6.57$, SD=0.65), which aligns with our design goal of making the video more ``informative''. Meanwhile, the improved immersion in video-consuming experience enabled by \textsc{Spica} was similarly well-received, evidenced by a mean score of 6.29 (SD= 0.83).

For both immersion and understanding, the mean scores for the \textsc{Spica} condition are higher (Understanding: $\mu=6.11, \sigma=0.84$; Immersion: $\mu=6.25, \sigma=0.86$) compared to the baseline condition (Understanding: $\mu=4.79, \sigma=2.10$; Immersion: $\mu=4.29, \sigma=1.84$), suggesting that participants rated \textsc{Spica} more favorably in both aspects. We use the Wilcoxon Signed-Rank test to analyze the significance between the rating scores for \textsc{Spica} and baseline in both metrics. For \textit{Understanding}, Z-score=2.12, $p=0.033<0.05$; for \textit{Immersion}, Z-score=2.00, $p=0.046<0.05$. The results showed there is a statistically significant difference in the ratings of both metrics between the \textsc{Spica} and baseline conditions. The result is demonstrated in Table~\ref{tab:wilcoxon}. \looseness=-1


\vspace{0.5em}
\noindent\textbf{{{Qualitative feedback regarding video understanding and immersion.}}} A majority of the participants complimented \textsc{Spica} for having more detailed insights and offering more options to explore while viewing videos. According to P3, \textit{``(\textsc{Spica} condition) gives more options for blind people, and allows for a better understanding of what is occurring in the videos.''} Specifically, P9 mentioned that \textsc{Spica} offers the liberty to \textit{``go back and explore the scenes,'' and ``provide the depth information to help better understand the video''}. P6 expressed that \textit{``the detailed information could fill in gaps that traditional audio descriptions miss, offering a richer viewing experience.''}

\begin{table}[htb]
\begin{tabular}{ccccc}
\hline
         & \multicolumn{2}{c}{\textbf{Understanding}} & \multicolumn{2}{c}{\textbf{Immersion}} \\ \hline
         & Mean             & SD            & Mean           & SD          \\
\textbf{\textsc{Spica}}    & 6.11*             & 0.84           & 6.25*          & 0.86         \\
\textbf{Baseline} & 4.79             & 2.10            & 4.29           & 1.84         \\ \hline
           
\end{tabular}
\caption{Users' ratings on their understanding and immersion for video consumption experience using \textsc{Spica} and Baseline. (*p<0.05)}
\label{tab:wilcoxon}
\end{table}

\subsection{{Effectiveness and User Perceptions of the System Features}}
We analyzed the participant feedback of each feature in \textsc{Spica} based on the post-study questionnaires and the qualitative analysis result. The analysis procedure is described in Section~\ref{sec:qualitative_method}.

\subsubsection{Temporal frame exploration}
Participants found the temporal frame exploration feature helped them develop a deeper connection with the storyline by allowing them to actively navigate through scenes per their preferences. For example, P1 values her ability \textit{``to go back and investigate the scene that I missed during the initial viewing.''} Similarly, P4 expressed that the feature gave her \textit{``a lot more control over my own viewing experience.''} enabling her to actively explore and navigate the narrative at her own pace. Participants also used the feature to skip scenes that were not interesting to them (P4, P5). Participants also expressed that navigating to the adjacent scenes instead of a fixed time interval made them feel \textit{confident} (P9, P11) on not missing important information in the video.

Concerns raised by participants about the frame exploration feature were mostly about narrative coherence and time cost. Some participants felt that frequent pauses of the video would affect the overall experience as they \textit{``did not want the original audio track to disappear for long.''} (P10). The participant also pointed out that \textsc{Spica} might be annoying when it paused at times when he only sought a little more details. Another participant (P13) pointed out the significant time the scene exploration process might take, stating that \textit{``You could easily spend an hour if you wished to delve into each and every frame for a 10-minute video.''} Others suggested that getting used to the new system and integrating it seamlessly into their viewing habits would \textit{``come with time''} (P12). 

Based on these observations, we found that allowing users to temporally pause the video and explore the timeline would help them capture more details and thus increase the overall understanding of the video; however, whether BLV users liked the pauses was a personal preference.

\subsubsection{Object exploration}
\label{sec:rq2_obj_exploration}
We found that 39.4\% (the percentage is calculated by dividing the number of frames that are explored in object level by the total number of frames where the video pauses at) of the paused frames, either due to the original extended AD or actively by the participants, were further explored by the participants at the object level. However, the percentage differs among videos, with the highest being 76.4\% (V3) and the lowest being 23.2\% (V2). This may be due to different topics and the quality of the original ADs. There was also a great variance in this measure among participants, suggesting diverse personal preferences. For example, P4 accessed object information on 92.3\% of the frames she explored, while the rate was only 13.4\% for P13.


Many participants found that the object exploration feature complements the frame exploration by enhancing their understanding of the frame/scene through object information. It helps to discern details such as facial expressions and outfits of speakers (P3, P4) and improves their spatial perception of where different objects are (P7). In Section~\ref{sec:mental_model} we discuss the detailed decision-making process of the participants about accessing object information.

\vspace{0.5em}
\noindent\textbf{{{Preferences between touch- and keyboard-based object exploration approaches.}}} {Among all participants, 6 participants used both input modalities (keyboard and touch) in the experiments. 4 out of 6 participants preferred to use keyboard-based exploration mainly because of its ease of use and easy transitions to other features in the system such as video playback control and frame exploration mechanism. For instance, P12 stated that: \textit{``As I mainly use the keyboard to use this system, I don't quite want to switch to other approaches like touch.''} Additionally, participants pointed out that keyboard exploration was more \textit{``deterministic''} (P9) so that they did not miss any objects in the frame. \looseness=-1}

{On the other hand, 2 out of 6 participants who preferred touch-based exploration mainly because of the spatial awareness that the approach supported. P10 explained that: \textit{``exploring using my fingers augments my perception towards the relative positions of different objects''}. Similarly, P11 expressed that: \textit{``I like when I touched a point at the screen and a spatial sound just coming from that direction...I felt it connected me with the scene in the video.''}}

\subsubsection{Object-specific spatial sound effects} 
Most participants found the retrieved object-specific sound effects useful. This is reflected in the post-questionnaire (Fig.~\ref{fig:post_quesionnaire}) where the mean score for the effectiveness of sound effects is 5.86 out of 7 (SD=1.51). Specifically, for the spatial effects added to these sound effects, the mean score is 5.21 out of 7 (SD=1.67). Participants agreed that the spatial sound effects were helpful in enhancing the viewer experience, particularly in immersion and spatial comprehension. They also noted that the accuracy and consistency of the retrieved sound effects are vital to their usefulness.

Participants expressed that the sound effects did not bring extra mental loads to their video viewing experience, instead, it \textit{``adds extra fun''}, and contributes to the immersion experience (P1). Additionally, it helps them develop a richer understanding of the scenes. For instance, P4 stated that the object description and the accompanying sound effects were especially useful when there was a complex scene with multiple elements to analyze, such as distinguishing between a bird, a car, and a person. In particular, P12 noted that the spatial effects were helpful in distinguishing the positioning of objects in a frame. P11 highlighted that spatial effects helped him \textit{``accurately placing them in space''} in his mind.

\vspace{0.5em}
\noindent\textbf{{Comparison with the original sound effect from the sounding object. }}We asked participants to compare how they felt about the retrieved sound effects to their feelings about the sound of the object in the original video soundtrack. 12 out of 14 participants expressed that the retrieved sound effect was even more helpful than the original sound effects in helping them to understand the spatial position of the objects and increase immersion while consuming the video. Noticeably, the retrieved sound would not confuse them, even though it was not from the original soundtrack; however, they need to be informed about the use of the sound retrieval mechanism.  For instance, P1 stated that: \textit{``I think when I first heard the noises of the different effects, it was a bit weird because it wasn't (from the original soundtrack). But once you explained, it all made sense.''} We learned that users could understand that the retrieved sound was used merely as a representation of the object and did not necessarily indicate the state of the object (e.g., a parked car can still have the ``zooming'' sound to represent it). Additionally, P7 noted that: \textit{``as some of the objects did not make sound, adding retrieved sound effects gave me a feeling to walk close to that object and observe it''}.

However, some participants reported feelings of distraction or confusion when the sound effect is not aligned with the actual objects. For instance, P14 reported that the sound effects could sometimes \textit{``mislead''} or \textit{``startle''} him when different objects under the same category have the same sound effect. For example, he felt confused \textit{``...whenever I touched a car, the sound effect was the same honking. This is strange because you can't have all the same cars parked on the street.''}


\subsubsection{High-contrast color mask} There were four participants who considered themselves \textit{``Legal blindness''} or \textit{``Blindness with some light/color perception''} and still had the ability to discern the high-contrast color mask of the object. All of them expressed a positive perspective on this feature. P6 mentioned that, with the mask, she could \textit{``easily locate and identify where the object is in the frame''}. Similarly, P14 confirmed the usefulness of this feature in facilitating faster localizing by saying: \textit{``Once it took all the color (of the other objects) away, it was a lot easier to find what I want.''}; however, he also expressed that the high-contrast color mask could not help him in \textit{``knowing exactly what the object is.''} This reflection is also echoed by P9 who still mainly relied on the audio clues to help consume the video content. However, he also noted that the use of color mask \textit{``doesn't require me to put extra attention... is a good addition when I was exploring the objects.''} Concluding from the participants' feedback, the high-contrast feature reduces users' difficulty in recognizing the objects without adding extra burden to them. However, its efficacy varies among users with different vision conditions.


\subsection{Triggers and Motivations of the Active Explorations from Users}
\label{sec:mental_model}

\begin{figure*}
    \centering
    \includegraphics[width=0.9\textwidth]{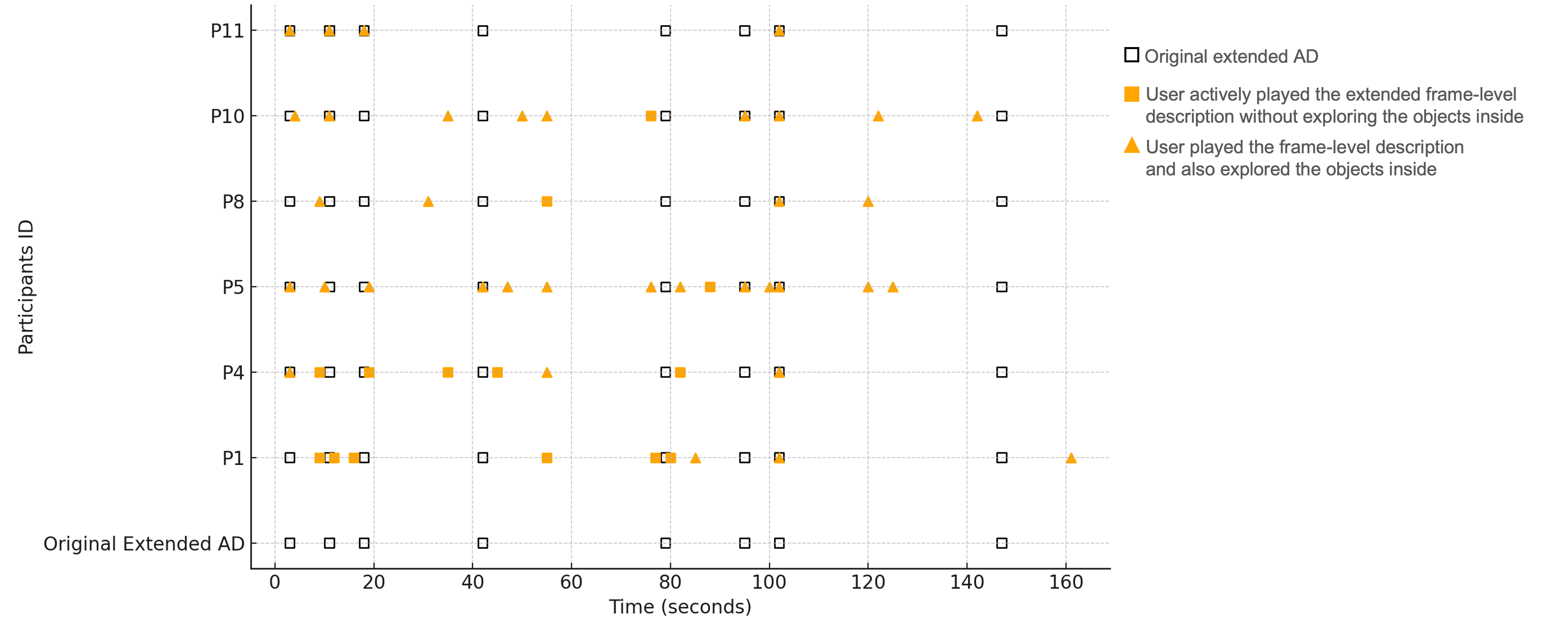}
    \caption{{User behaviors of a sample video (V6). A total of 6 participants (P1, P4, P5, P8, P10, and P11) experienced this video using \textsc{Spica}}}
    \label{fig:pause}
    \Description[Participants showed different exploration patterns.]{Different participants showed various exploration patterns. 61.4 percent of the active frame exploration happened with an interval of ±5 seconds of a frame with an original AD.}
\end{figure*}

We analyzed the trigger events and motivations of the user when engaging different interactive features of \textsc{Spica}. We focused on their active pauses, frame exploration, and object exploration.


\vspace{0.5em}
\noindent\textbf{F1: Original ADs have an anchor effect on users' active video pausing and frame exploration.} We found that participants tend to explore adjacent frames of a frame that has an original audio description. According to the video annotation result, 61.4\% of the active frame exploration happened with an interval of $\pm$5 seconds of a frame with an original AD. Furthermore, we found that of all object explorations around the original AD, 37.4\% happened exactly at the frame where the original AD exists, showing that users often wanted more details of the frame where an original AD was played. To illustrate this distribution, we sampled a video (V6 from Table~\ref{tab:videos}) and plotted active behaviors of the users in Fig.~\ref{fig:pause}. The result is also reflected in the interview, where P5 stated that: \textit{``obviously the automatically played (original) audio description gets more information for me compared with just hearing the sound from the video, but it also brings more curiosity to me so that I would want to pause and do some explorations around.''} Similarly, P10 mentioned that: \textit{``I used to watch a lot of video content with (original) audio descriptions every day, and those helped a lot. So I would assume those are frames that are worth exploring, given I have the power to do so (using \textsc{Spica}).''} Noticeably, as the video is already paused at the original extended AD, users may be more inclined to explore the scene further, as this action does not demand a considerable extra effort or significantly heighten the feeling of interruption.

\vspace{0.5em}
\noindent{\textbf{F2: Users pause or temporarily explore the video for reflecting, validating, and customizing their content consumption experience.} For reflecting,} P4 mentioned that when she perceived multiple events happening in a frame (e.g. several conversations happened simultaneously, various types of sounds existed in the video...), she would wonder what the characters were doing and how those actions were associated with each other. At this time, pausing allows her to digest and reflect on the information for a deeper understanding of the content. {For validating,} P9 noted that he would navigate to the previous frame when he was \textit{``curious on what leads to the current description''}. He explained that: \textit{``(at a frame) the description was: 'the two casts hugged.' I was like, what happened before they hugged? Did they have any eye contact or something?''} Similarly, P8 detailed this experience and said that: \textit{``from the environmental sounds (from the video), I know the scene switches scenes very quickly, for example, they may show some people at home, and then they may show some other group of friends that are hanging out at a bar. At this time I feel I cannot learn anything from such switching.''} Therefore, going back to a prior frame ensures that the user can have a coherent and continuous narrative for developing a complete storyline. There are also cases where users pause and explore the scene when \textit{``the (original) audio descriptions disappear for so long, while the audio in the raw video also keeps unchanged.''}(P12). {For customizing the consumption experience}, P3 expressed that when the video content was less engaging---usually when the original ADs were absent for a long period and the original soundtrack barely changed---she would fast-forward to the latter frames. When she did so, she appreciated that \textsc{Spica} reads out scene descriptions immediately when she navigated, which helped her to customize their viewing experience on more relevant and engaging content. 



\vspace{0.5em}
\noindent\textbf{F3: The granularity and quality of the frame-level description affect users' decisions on whether to access object information further. }
In general, we found if the frame-level description is already in details, users tend to skip object exploration. For instance, P1 noted that \textit{``I feel like I don't need to explore the objects and understand it more because the frame description felt self-explanatory.''} Similarly, P9 echoed that \textit{``If the frame description has quite a bit of detail in it, I don't need to explore (the objects) further.''} On the contrary, if the frame-level description is less informative, users would also skip object exploration. For example, P8 explained that: \textit{``The frame didn't really seem to be super detailed, so I assumed the scene did not have many interesting things to explore.''}

We also found that users would like to explore more on the frames that originally have extended ADs, compared with those they actively paused. Quantitatively, we found users took object exploration on 43.8\% of the frames that have native extended ADs, while the percentage is 33.2\% for the frames they actively paused. This could be because the original ADs are often positioned in more crucial places that are worth exploring. It is also possible that users are more inclined to explore further when the video is already paused.

The results also demonstrate a common pattern where the frame description gave the user a high-level overview of the scene, and the object descriptions complement it with more details. For example, P14 mentioned: \textit{``Sometimes I could tell if this scene has changed from the raw video sound, usually when it was a big (complicated) scene, I don't get enough explanations until I delve into the frames to scan through all the objects.''} Similarly, P7 stated that: \textit{``The frame description told me there were cars, stop signs and many other things on the street, which made me curious on where they were and what they looked like.''} 


\vspace{0.5em}
\noindent{\textbf{F4: Users often explore individual objects when they hear new sounds from the original video or they hear particular objects mentioned in the ADs.}} {One of the most common user information needs is identifying the object or activity causing a sound. Analyzing video-based behavior coding revealed that 14.9\% of frames where users explored objects immediately followed the introduction of new sounds.} For example, P5 mentioned that: \textit{``When I heard the car's engine sound, I would like to know where that car was and why it was there.''}. Similarly, P4 mentioned that: \textit{``When I heard a lot of noisy conversations happening around, I'd kind of want to know what are they doing?''}; \textit{``when I heard there was a new woman role coming in, I would like to learn what she looks like.''}

We also found that users often have specific information needs about the characteristics of the object. For example, several participants expressed their need to explore the details of the speakers in the video. P1 stated that: \textit{``I am always curious to know what clothes people are wearing.''} Similarly, P13 was particularly interested in the facial expressions of speakers, she said that: \textit{``I heard the description that a man is sitting in front of the laptop, I’m wondering was it like a stressed man, flustered man or a happy man?''}. Also, P14 mentioned that: \textit{``I stopped (and explored) it... where the family was standing at the table introducing themselves, so I wanted to stop it and go through with the individuals to learn more details about their clothing and stuff.''} However, in contrast, such detailed descriptions were often left out in the original AD or frame-level descriptions. Some participants tend to explore the surroundings thoroughly when they hear specific sounds, such as a door closing or a phone ringing. P8 expressed that such explorations \textit{``help me to actually immerse in the environment, it also makes me feel more confident in understanding their conversations (in the video) by giving more context.''} 

Our result shows that although original audio descriptions are often designed to describe the scene in important scene changes, it would be useful to allow users to access the scene layout (including the ``less important'' ones) as they need.




\subsection{Handling AI Errors, Inconsistencies, and Discrepancies}
\label{sec:error_handling}
The object recognition and caption generation pipelines used in \textsc{Spica}, while generally performing well, still make errors sometimes as shown in Section~\ref{sec:relatedwork_ad_gen}. It is vital that BLV users minimize the impact of inaccurately generated descriptions, as they can be misleading. Our results suggest that BLV users could address some of the errors in the generated captions by examining the context with the help of \textsc{Spica}.

During our user study with BLV participants, we found that there were several times when participants encountered and reported inaccuracies in the predicted results. By interviewing and discussing with the participants, we found that inconsistencies between the audio narrative and frame or object descriptions enabled users to spot errors. For example, P5 shared, \textit{``(At first) I felt really confused when the description told me there was a second person in the scene, but I only heard one person's sound.''} Yet, by examining adjacent frames, he determined the likely error: \textit{``Then I explore other frames around that frame, turning out the descriptions to the neighboring frames only depicted one person. Then I thought I was deterministic enough to say the description to that frame was wrong.''} 

Likewise, P13 articulated her ability to navigate occasional wrong predictions by investigating other objects in the scene: \textit{``At least I can find the object that is mentioned in the frame or sounded in the video itself, I think I am able to comprehend the broader picture presented by the video... So I’m willing to take the wrong predictions.''} She also gave another concrete example of how she inferred that the prediction was wrong: \textit{``When I heard there was an accordion as I explored. I feel that's probably not, because the scene was a kitchen, maybe it was a cutting board or a washcloth I guess.''}

There were also cases where the user could not correctly handle the errors. This happened when the object detector detected an object that was not mentioned in the frame-level description and did not make sound in the video, or the scene changed frequently while the original ADs were absent. In this case, participants usually skipped the part and continued their video consumption experience: \textit{``I understand this is a common case in using AI-assisted tools, so I just choose to omit the information the system provides and stick to the original video sound and audio descriptions.'' (P9)}


%% file: 6-DiscussionFutureWorkConclusion.tex
\section{Discussion}

\subsection{Tradeoffs of Information and Cognitive Load}
{BLV users have varied preferences on information granularity and understanding ability; therefore, systems should support the user's agency in customizing the types, amount, forms, and modalities of information they receive.}

Recent research~\cite{fan2023improving} has demonstrated the importance of reducing cognitive load while preserving key information to improve the accessibility of the multimedia content consumption experience (e.g., screen-shared presentations). In the context of video consumption, detailed audio descriptions help users understand video content; however, increasing the level of detail also raises the cognitive burden. While sighted individuals simultaneously process visual and auditory information, BLV users predominantly depend on auditory information. This reliance can increase the cognitive strain when consuming content. Feedback at the moment of a \textit{trigger event}, as discussed in Section~\ref{sec:mental_model}, is often preferred by users, but conventional Audio Descriptions (AD) often omit or delay such feedback, adding to mental load.

\textsc{Spica} tackles this challenge with \textit{layered} captions. The base layer is the original AD, but users can opt to access extra information through automatically generated object captions.

\subsection{Augmenting Audio Experience}
\label{sec:discuss_audio}

In addition to providing verbal descriptions of scenes and objects to users, \textsc{Spica} enhances auditory experiences by adding relevant sound effects linked to the objects, thus creating a deeper sense of immersion when viewing videos. Importantly, \textsc{Spica} uses the calculated 3D object position to generate spatialized sound effects, enhancing the user's perception of the object's location. Feedback from our study reveals that users appreciate this feature, as auditory cues naturally augment video viewing.

Most of the participants deemed the sound effects introduced beneficial for immersion. This observation underscores the value of integrating additional information, even if not directly extracted from the video, to elevate the viewing experience for the BLV audience. Participant P9 emphasized the significance of contextually appropriate sound effects, suggesting: \textit{``(Instead of using the honk sound only) it would be more interesting if it overlays the background sound, for example, the actual sound on the street.''}  Future research can explore refining the alignment of sound effects with objects and their environment, potentially using advanced retrieval or sound synthesis models like~\cite{agostinelli2023musiclm}

Furthermore, the spatial audio technique in \textsc{Spica} confirms the power of spatial elements to improve the user experience. While \textsc{Spica} currently employs a \textit{static} spatial audio effect to indicate object positioning, an intriguing expansion would incorporate temporal cues in spatial sound. This would allow users to discern object movements within a stationary visual scene, mirroring Participant P5's feedback: \textit{``It would be even more immersive if I can hear the car's sound moving in the space.''}

\subsection{Multisensory Video Consumption}
\subsubsection{{Employing tactile and gesture feedback}}
In \textsc{Spica}, we enable users to interact with a visual scene through touch devices, fostering an intuitive understanding of the object's 2D position and the spatial relationships between various objects in the frame. This insight paves the way for future tools that employ additional sensory channels to enhance the experiences of BLV users.  One promising direction could be to integrate haptic feedback during object-level exploration, allowing users to understand an object's shape and texture more deeply. Recent research has also demonstrated the potential to computationally transform 2D audio or visual data into 3D environments~\cite{ningmimosa,zhang2019predicting,zhang2023peanut}.  Such advances may be used to enable innovative interactive methods in VR, encouraging users to employ various gestural interactions for video exploration.

\subsubsection{{Touch-based object exploration versus voice-based approach}}
\label{sec:touch_vs_voice}
{\textsc{Spica} supports both keyboard-based and touch-based interaction techniques for object exploration. Based on multimodal interaction theory~\cite{multimodal,multimodal2,wolff1998acting}, the touch-based approach integrates the touch and auditory perceptions of the BLV participants, enhancing their overall experience. Additionally, this approach intuitively develops a mental mapping of the object's spatial position layout, specifically, the position where the user's finger points is the relative place of the object to the scene, which is an intuitive way to augment the spatial awareness of the BLV participants. However, users may miss out on objects if they do not thoroughly explore the scene. For complex scenes with numerous objects, this strategy might also be overwhelming for BLV users.}

{An alternative way is to design a voice-based object exploration strategy, where users can simply speak out what object they want to learn about and the system can give an audio description of that object accordingly. Using an agent, the strategy allows users to get immediate feedback on the objects they want to explore~\cite{voice1,voice2,voice3}. However, this strategy lacks discoverability for BLV users. Additionally, the agent can cause user frustration or misinformation associated with inaccuracies in voice recognition and intent classification~\cite{geno,sugilite,li2020multi,li2018appinite}.} \looseness=-1

\subsubsection{{Augmenting visual modality to assist people with diverse types of visual impairments}}
{Meanwhile, we find that careful use of visual modality can still benefit low-vision users with some access to visual information.} In \textsc{Spica}, we use the high-contrast color mask to highlight the selected object in the visual scene. Going further, it can be valuable and effective to design accessible video consumption tools that accommodate varying degrees of visual impairments. For example, for users with color vision deficiencies such as Daltonism, we can design an adaptive system to provide alternative color schemes for different portions of the visual scene based on the user's vision condition. Furthermore, there is extensive research on the use of color cues to improve the learning of color-blind users~\cite{richardson2014color,wurm1993color}. In the video consumption domain, it would also be beneficial to develop intuitive approaches to use color cues as indicators of temporal variation. For instance, using a gradual change in color intensity to signal the transitions in the states of objects.

\section{Limitations and future work}
\subsection{Adapting to Individualized Video Exploration Preferences}
\label{sec:future_adaptive}
{Our user study demonstrated a wide range of user preferences regarding narration styles, detail levels in descriptions, and exploration methods among BLV users. A key focus of our future work will be to enhance \textsc{Spica}'s adaptability to accommodate diverse user needs and preferences.}

Within the current version of \textsc{Spica}, we have a predefined narration style for descriptions. For instance, object-level descriptions begin with the object index, then detail the 2D position in the frame, and finally describe the object. Although suitable for many, some participants (e.g., P5, P8) found this too verbose, preferring to rely solely on spatial sound cues for object location.

Furthermore, users' preferences for description granularity varied. Some wished for exhaustive frame-level narratives that incorporate in-depth object details, while others favored concise summaries for a quick scene overview without disrupting their viewing experience. A notable suggestion was to refine frame-level descriptions to fit seamlessly between sentences in the original video, especially for those who favor inline ADs over extended ones.

Regarding object exploration, participants sought more versatile and customizable methods beyond what \textsc{Spica} currently offers. Specifically, some users prefer to \textit{``hear all the object descriptions to be read out at once.''} (P11). Some other users, preferring keyboard-based exploration, suggested the ability to jump directly to an object using its index rather than navigating sequentially.

\subsection{Improving the Description Generation Pipeline}

In \textsc{Spica}, {we primarily leveraged off-the-shelf state-of-the-art ML models to implement a new description generation pipeline, whose performance has been validated by an offline technical evaluation. Building upon the insights from our user study, we have identified additional opportunities for future work in model development to better align this pipeline with the actual needs of BLV users.}

{As discussed in Section~\ref{sec:error_handling}, there are disconnections when the original sound of the video does not seamlessly integrate with our frame or object-level descriptions. To address this, we intend to integrate models such as~\cite{Tian_2018_ECCV,Tian_2020_eccv,tian2021cyclic} that combine visual and audio information. In addition, we are considering the implementation of advanced vision-language models such as BLIP~\cite{li2022blip} and GPT-4V\footnote{\url{https://platform.openai.com/docs/guides/vision}} to enhance the quality of the description. These improvements could pave the way for the integration of interactive visual question answering (VQA), further helping BLV users understand the video content at the frame and object levels.}

\subsection{{Improving Utility for Longer Videos and  Group Watching Experience}}

Although \textsc{Spica} demonstrated promising results in the user study, we only selected short videos in the study due to time constraints. Its efficacy on longer content, such as full TV episodes or animations, has not been fully tested. {Additionally, in the current version of the system, we focus on improving the individual consuming experience for BLV participants by enabling personalized consuming experiences. However, the personalized nature of \textsc{Spica} may hinder its use in a group-watching setting, as the video progress depends on the user's exploration preferences, and therefore does not synchronize within a group.} 

As a next step, our aim is to launch a field study that will allow BLV users to incorporate \textsc{Spica} into their daily routines to consume a large variety of videos in their actual daily context. This deployment will validate the ecological validity of \textsc{Spica} and provide valuable insight into how BLV users interact with \textsc{Spica} within its intended context of use. {Findings of this future study can help us address these practical issues regarding the user experience of \textsc{Spica} that are related to specific usage contexts.} In the end, we plan to release \textsc{Spica} to the public to promote the positive broader impacts of our work in the BLV community.

\section{Conclusion}
In this paper, we presented \textsc{Spica}, an AI-powered video consumption tool that enables BLV users to interactively explore video content through augmented audio descriptions. \textsc{Spica} features a layered interaction strategy that allows users to explore the visual scene on both time and spatial levels, and further augment their object exploration experience through spatialized sound effects and high-contrast color masks. A within-subjects user study demonstrates that \textsc{Spica} is able to improve BLV users' understanding and immersion in the video content. Our findings also provide design implications for interaction-based accessible video consumption tools in the future.